\documentclass[shortnote,twocolumn]{jpsj2} 
\begin{document}
\title{Anomalous Anisotropic Magnetoresistance in Heavy-Fermion PrFe$_{4}$P$_{12}$}
\author{Hitoshi \textsc{Sugawara}\thanks{E-mail address: sugawara@ias.tokushima-u.ac.jp}, Eiichi \textsc{Kuramochi}$^1$, Takahiro \textsc{Namiki}$^{1}$, Tatsuma D. \textsc{Matsuda}$^{1}$\thanks{Present address: Advanced Science Research Center, Japan Atomic Energy Agency, Tokai, Ibaraki 319-1195}, Yuji \textsc{Aoki}$^1$, and Hideyuki \textsc{Sato}$^1$}
\inst{
Faculty of Integrated Arts and Sciences, The University of Tokushima, Tokushima 770-8502, Japan\\
$^{1}$Department of Physics, Tokyo Metropolitan University, Minami-Ohsawa, Hachioji, Tokyo 192-0397, Japan
}
\kword{filled-skutterudite, PrFe$_4$P$_{12}$, heavy-fermion magnetoresistance, anisotropy}
\maketitle

PrFe$_{4}$P$_{12}$ with a filled-skutterudite structure (${\it Im}$\={3}) has attracted much attention because of an apparent Kondo effect, an unusual nonmagnetic ordered state (A-phase) below $T_{\rm A}=6.5$~K, and a heavy-fermiom (HF) behavior at high fields where the A-phase is suppressed.~\cite{Sato_PRB,Aoki_PRB,Sugawara_PRB,Aoki_JPSJ} The $c-f$ hybridization and multipolar interaction are believed to play key roles for the anomalous properties. 
One of the recent fascinating findings is the discovery of a high-field ordered phase (B-phase), which is observed only in limited field directions around $H\|[111]$.~\cite{Tayama} 
At high fields, we have found a highly anisotropic mass enhancement both in the dHvA and specific heat experiments; the mass enhancement is largest for $H\|[111]$ in the three principal directions.~\cite{Sugawara_PRB,Namiki} In addition to the anisotropy, the mass enhancement is suppressed with increasing magnetic field. In preliminary experiments on magnetoresistance (MR) below 8~T,~\cite{Kuramochi} we have found a unique field angle dependence of MR, which shows a sharp peak around $H\|[111]$. It is naturally expected that the B-phase is related to the anisotropic mass enhancement and also the anisotropic MR. In this paper, we report on MR measurements with an emphasis on the magnetic anisotropy at the high fields of PrFe$_4$P$_{12}$ to deepen our understanding of unusual properties.

Single crystals, whose qualities are basically the same as those used for dHvA experiments, were prepared by a Sn flux method.~\cite{Sugawara_PRB} Magnetoresistance was measured by a conventional DC four-probe method in a top-loading $^{3}$He cryostat with a 16~T superconducting magnet. All measurements were performed using a transverse geometry, as shown in Fig.~3(a). 

\begin{figure}[tp!]
\begin{center}\leavevmode
\includegraphics[width=0.7\linewidth]{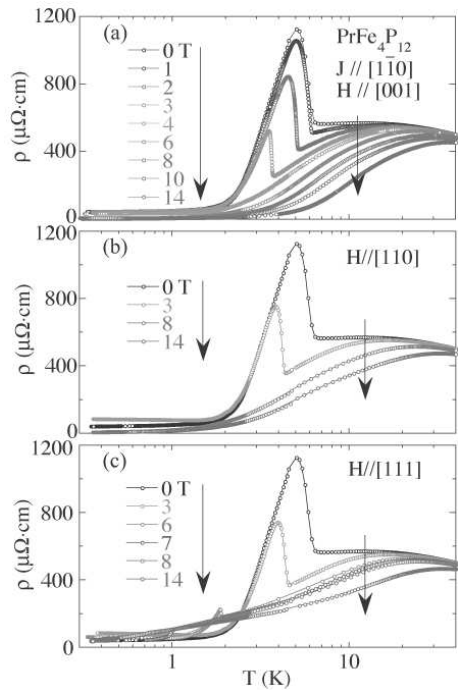}
\caption{(Color online) Temperature dependence of the electrical resistivity at selected fields for (a) $H\|[001]$, (b) $H\|[110]$, and (c) $H\|[111]$ in PrFe$_4$P$_{12}$.}
\label{R_T}
\end{center}
\end{figure}

Figure~\ref{R_T} shows the temperature $T$ dependence of the electrical resisitivity $\rho(T)$ below 40~K at selected magnetic fields along $H\|[001]$, $H\|[110]$, and $H\|[111]$.
$\rho(T)$ at zero field is basically the same as that reported previously; it increases below $\sim200$~K almost logarithmically with decreasing $T$ (not shown here) and shows a faint maximum at $\sim13$~K.~\cite{Sato_PRB} The sharp upturn at around $T_{\rm A}=6.5$~K reflects a reduction in the number of carriers due to Fermi surface (FS) reconstruction.~\cite{Harima,Sugawara_PRB}  With increasing $H$,  $T_{\rm A}$ shifts to lower temperatures, and the A-phase is completely suppressed above the metamagnetic transition field $H_{\rm M}$, above which the HF state has been confirmed both in the specific heat and dHvA experiments.~\cite{Aoki_PRB,Sugawara_PRB} Also, on $\rho(T)$ under the magnetic fields for $H\|[001]$,~\cite{Sugawara_PRB,Aoki_JPSJ} we have found the large coefficient $A$, which follows the Kadowaki-Woods relation.~\cite{Kadowaki} We note here that the low-temperature resistivity slightly increases at 3~T for all field directions. Recent neuron experiments revealed that the field-induced antiferromagnetic (AFM) moment increases with increasing $H$ in the A-phase.~\cite{Hao} The increase in magnetic scattering intensity due to the induced AFM moment may be the origin of the resistivity increase in the A-phase.
 
\begin{figure}[tp!]
\begin{center}\leavevmode
\includegraphics[width=0.95\linewidth]{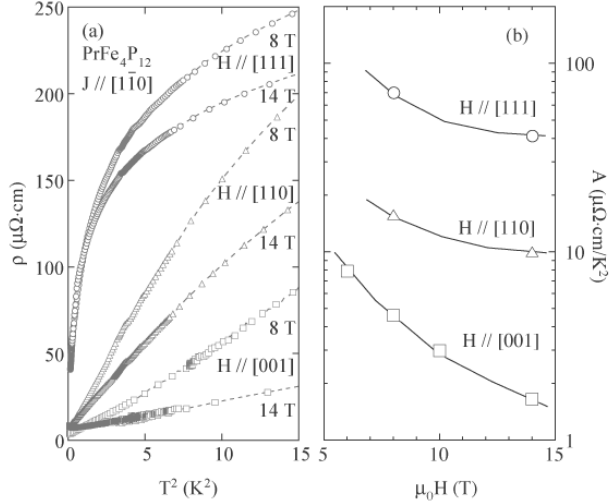}
\caption{(Color online) (a) Comparison of the $T^{2}$ dependence of the electrical resistivity at 8 and 14~T for fields along $H\|[111]$, $H\|[110]$, and $H\|[001]$ in PrFe$_4$P$_{12}$. (b) Field dependence of coefficient $A$ of $\rho=\rho_{\rm 0}+AT^{2}$ in PrFe$_4$P$_{12}$. The solid lines in Fig. 2(b) are guides to the eye.}
\label{A_H}
\end{center}
\end{figure}

In this experiment, as shown in Figs.~\ref{R_T}(b) and \ref{R_T}(c),  we have further measured $\rho(T)$ under the magnetic fields for $H\|[110]$ and $H\|[111]$, and determined the anisotropy of $A$ and its field variation up to 14~T, as shown in Fig.~\ref{A_H}. 
Both for $H\|[110]$ and $H\|[001]$, as shown Fig.~\ref{A_H}(a), $\rho(T)$ clearly exhibits $T^{2}$ dependence. $A$ at 14~T for $H\|[110]$ estimated below $\sim2$~K is about 6 times larger than that for $H\|[001]$. The Fermi-lquid behavior, however, evidently breaks down for $H\|[111]$, where $\rho(T)$ exhibits $T^{0.5}$ dependence for a wider temperature range below $\sim1.8$~K and can be fitted by $T^{2}$ only for limited temperature ranges just above the transition temperature of the B-phase.~\cite{Tayama,comment} The estimated $A$ at 14~T for $H\|[111]$ is about 25 times larger than that for $H\|[001]$. As shown in Fig.~\ref{A_H}(b), we have also confirmed that $A$ is suppressed by the magnetic field for $H\|[110]$ and $H\|[111]$. Such a suppression is consistent with the results of dHvA and specific heat experiments, in which the effective mass is strongly suppressed by the magnetic field.~\cite{Sugawara_PRB,Namiki} Here, we note that the field variation is less sensitive for the fields along $H\|[110]$ and $H\|[111]$ than that for $H\|[001]$ within the experimental field range; the $A$ ratios between 8 and 14~T are $A_{\rm 8T}/A_{\rm 14T}=2.8$ for $H\|[001]$, $A_{\rm 8T}/A_{\rm 14T}=1.6$ for $H\|[110]$, and $A_{\rm 8T}/A_{\rm 14T}=1.7$ for $H\|[111]$, though the origin is not clear at this stage.

\begin{figure}[tp!]
\begin{center}\leavevmode
\includegraphics[width=0.65\linewidth]{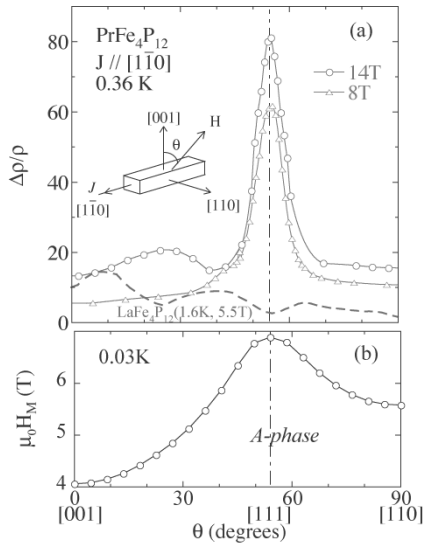}
\caption{(Color online) Angular dependence of (a) the magnetoresistance and (b) metamagnetic transition fields in PrFe$_4$P$_{12}$. The dashed line in Fig. 3(a) indicates the angular dependence of magnetoresistance in LaFe$_{4}$P$_{12}$.}
\label{AngDepMR}
\end{center}
\end{figure}

Figure~\ref{AngDepMR}(a) shows the angular ($\theta$) dependence of magnetoresistance $\Delta\rho/\rho=[\rho(H)-\rho(0)]/\rho(0)$ at 0.36~K under the magnetic fields 8 and 14~T, where the resistivity at zero field $\rho(0)$, which should be the residual resistivity of an unordered state, is estimated to be $\rho(0)=0.57\mu\Omega\cdot{\rm cm}$ from the $H^{2}$ dependence of $\rho(H)$ at high fields for $H\|[001]$ in $H\rightarrow0$~T.
The configuration of the magnetic field $H$ and the current $J$ directions are shown in the inset of this figure. The angular dependence at 8~T shows a sharp peak at around $H\|[111]$ with a half width of about $9^\circ$, as reported perviously.~\cite{Kuramochi}  This peak structure almost remains unchanged even at 14~T except that the absolute value of $\Delta\rho/\rho$ is enhanced about 1.3 times. On the contrary, a broad peak structure appears at around $26^\circ$.  
For comparison, we show the angular dependence of $\Delta\rho/\rho$ in LaFe$_{4}$P$_{12}$,~\cite{Sugawara_JPSJ} which is reasonably explained by the topology of the Fermi surface (FS). Evidently, the angular dependence of $\Delta\rho/\rho$ in PrFe$_{4}$P$_{12}$ is different from that in LaFe$_{4}$P$_{12}$, suggesting that the angular dependence cannot be explained by the topology of FS simply assuming the trivalent state of Pr ions in PrFe$_{4}$P$_{12}$. Therefore, such an anisotropy of $\Delta\rho/\rho$ may be ascribed to the magnetic origin, since the angular dependence of $\Delta\rho/\rho$ is clearly correlated with the anisotropy of magnetization and the metamagnetic transition field $H_{\rm M}$.~\cite{Aoki_PRB} For  comparison, the angular dependence of $H_{\rm M}$ determined in the dHvA experiments is shown in Fig.~\ref{AngDepMR}(b).~\cite{Sugawara_PRB} 

The nature of low-energy crystal field (CEF) levels is essential to low-temperature physical properties, such as magnetic anisotropy. Recently, Kiss and Kuramoto have proposed a $\Gamma_{1}$-$\Gamma_{4}^{(1)}$ scheme for the CEF levels,~\cite{Kiss} which is considered to be the most promising model for accounting for the high-field ordered state (B-phase). In this model, level crossing with increasing magnetic field is only possible for $H\|[111]$, which gives rise to the field-induced phase, as observed in PrOs$_{4}$Sb$_{12}$.~\cite{Kohgi} Using this model, they also calculated the angular dependence of the resistivity, which qualitatively explains the results of the present experiments. Very recently, on the other hand, they reported that the phenomenological theory of a scalar electronic order with a $\Gamma_{1g}$-type symmetry can reproduce magnetic properties, such as the isotropic magnetic susceptibility in the A-phase, the angular dependence of the transition temperature $T_{\rm A}$ under a magnetic field, and the splitting pattern of $^{31}$P nuclear magnetic resonance (NMR) spectra, ~\cite{Kiss2008} although the microscopic nature of the scalar-order parameter is still an open question. Further studies are necessary to clarify the ordered states, which is also important for understanding the high-field anisotropic HF state.

In summary, we have investigated the anisotropy of the magnetoresistance in the Pr-based HF compound PrFe$_4$P$_{12}$. The large anisotropy of effective mass and its strong field dependence have been confirmed by resistivity measurements. Particularly for $H\|[111]$, where the effective mass is most strongly enhanced, the non-Fermi liquid behavior has been observed. Also, we have found the angular dependence of the magnetoresistance sharply enhanced at $H\|[111]$, which is evidently correlated with both the non-Fermi liquid behavior and the high-field ordered state (B-phase). 
 
This work was supported by a Grant-in-Aid for Scientific Research Priority Area Skutterudite (Nos. 15072204 and 15072206) of MEXT, Japan.


\begin{thebibliography}{99}
\bibitem{comment}We have found the anomalies of resistivity corresponding to the B-phase transitions at  $T_{\rm B}=0.56$~K at 8~T and $T_{\rm B}=0.75$~K at 14~T (not shown).
\end{thebibliography}
\end{document}